\documentclass[english,aip, twocolumn, superscriptaddress]{revtex4}
\usepackage[LGR,T1]{fontenc}
\usepackage[latin9]{inputenc}
\usepackage{color}
\usepackage{ifsym}
\usepackage{amstext}
\usepackage{amsmath}
\usepackage{graphicx}
\usepackage{epsfig}
\usepackage{epstopdf}
\usepackage[sort&compress]{natbib}

\makeatletter

\@ifundefined{textcolor}{}
{%
 \definecolor{BLACK}{gray}{0}
 \definecolor{WHITE}{gray}{1}
 \definecolor{RED}{rgb}{1,0,0}
 \definecolor{GREEN}{rgb}{0,1,0}
 \definecolor{BLUE}{rgb}{0,0,1}
 \definecolor{CYAN}{cmyk}{1,0,0,0}
 \definecolor{MAGENTA}{cmyk}{0,1,0,0}
 \definecolor{YELLOW}{cmyk}{0,0,1,0}
}

\makeatother

\usepackage{babel}
\begin{document}

\title{Universal Conductance Fluctuations in Electrolyte-Gated SrTiO$_3$
Nanostructures}

\author{Sam W. Stanwyck}

\affiliation{Dept. of Applied Physics, Stanford University}

\author{P. Gallagher}

\author{J.~R. Williams}

\author{David Goldhaber-Gordon}

\affiliation{Dept. of Physics, Stanford University}

\date{\today}
\begin{abstract}
We report low-temperature magnetoconductance measurements of a patterned two-dimensional
electron system (2DES) at the surface of strontium titanate, gated
by an ionic liquid electrolyte. We observe universal conductance fluctuations,
a signature of phase-coherent transport in mesoscopic devices. From the universal
conductance fluctuations we extract an electron dephasing rate linear in temperature, characteristic of
electron-electron interaction in a disordered conductor. Furthermore, the dephasing rate has a temperature-independent offset, suggestive of unscreened local magnetic moments in the sample. 
\end{abstract}
\maketitle

Strontium titanate (STO) has been the subject of a resurgence of experimental interest, mainly because of the recent ability to create STO-based materials with electrons confined in one and two spatial dimensions\cite{hwang_emergent_2012-1}. STO is also known to host a wide variety of electronic ground states determined by the density of electrons. These two aspects combine to create a promising system for studying nanoscale electronics where the interactions between electrons can be controlled simply by a gate voltage. Two-dimensional electron systems in STO have primarily been created in three different ways: growing a polar overlayer, usually lanthanum aluminate(LAO), to create a heterointerface\cite{ohtomo_high-mobility_2004}, $\delta$-doping with Nb or La\cite{kozuka_two-dimensional_2009, jalan_two-dimensional_2010}, or with an electrolyte gate in an electric double-layer transistor (EDLT) configuration\cite{ueno_electric-field-induced_2008-1, lee_phase_2011,  lee_electrolyte_2011}. Low-temperature transport in these systems has revealed a wide variety of electronic ground states. Undoped STO is an insulator, but at $\sim$1x10$^{13}$ cm$^{-2}$ the system becomes a metal. At higher densities ($\sim$3x10$^{13}$ cm$^{-2}$) -- still much lower than typical BCS superconductors --  the system becomes a two-dimensional superconductor. Early evidence suggests that the entrance into a superconducting state is a Berezinskii-Kosterlitz-Thouless transition\cite{reyren_superconducting_2007}. At even higher densities, a ferromagnetic phase appears to exist\cite{brinkman_magnetic_2007, moetakef_carrier-controlled_2012} and may even coexist with superconductivity\cite{dikin_coexistence_2011, li_coexistence_2011, bert_direct_2011}.

The handful of experiments on the quantum transport properties of two-dimensional electrons in STO have thus far focused on locally patterned LAO/STO, and have produced several interesting observations. The Rashba spin-orbit interaction -- extracted from weak antilocalization measurements -- is relatively large and tunable by application of a gate voltage. Further increases of the spin-orbit strength have been linked to quantum critical points in the system\cite{caviglia_tunable_2010, ben_shalom_tuning_2010}. Universal conductance fluctuations (UCF) corresponding to phase coherence lengths of several microns have recently been observed in LAO/STO microstructures\cite{rakhmilevitch_phase_2010}. Recently, conductive atomic force microscope tips have been used to create nanoscale conduction paths at the LAO/STO interface\cite{xie_charge_2010, cheng_sketched_2011}.

In contrast, in electrolyte-gated STO some of the most basic properties of the quantum transport have yet to be investigated. In this work, we apply the EDLT technique to an undoped STO sample with nanopatterned metallic gates to study STO 2DESs with lateral confinement on the hundred-nanometer scale. At this length scale, UCF dominate the magnetoconductance, as demonstrated in LAO/STO samples\cite{rakhmilevitch_phase_2010, stornaiuolo_-plane_2012}. We analyze the UCF, extracting a phase coherence length $l_\phi$ as a function of temperature. The phase coherence length we separately extract from weak antilocalization agrees with that determined from UCF. The dephasing rate $\tau_\phi^{-1}$ depends linearly on temperature, suggesting that electron-electron interactions are one main source of low-temperature dephasing. However, extrapolation to zero temperature yields a nonzero $\tau_\phi^{-1}$, implying that electron-electron interaction is not the sole source of dephasing. We argue that interaction with localized magnetic moments could be responsible for the temperature-independent contribution to the dephasing rate.

Our sample consists of a 300 nm-wide Hall bar defined by nanopatterned gates within a larger STO Hall bar (Fig. 1a). The substrate is an undoped STO crystal that has been chemically treated and annealed to achieve nominal TiO$_2$ surface termination(Shinkosha Ltd, Japan). We pattern Titanium/Gold ohmic contacts, as in previous work, by photolithography, argon ion milling, and electron-beam evaporation\cite{lee_electrolyte_2011}. We define the nanoscale gates via electron-beam lithography; spincoating a conductive polymer (ESPACER 300Z) on top of the PMMA is required to reduce sample charging. Following PMMA development, we grow 10 nm of an alumina dielectric via atomic layer deposition, evaporate 5 nm titanium/30 nm gold, and perform liftoff in acetone. Finally, we pattern a layer of hard-baked photoresist with an opening to delineate the larger Hall bar and another opening for a coplanar gate. 

\begin{figure}
\label{fig:device}
\includegraphics[scale=0.6]{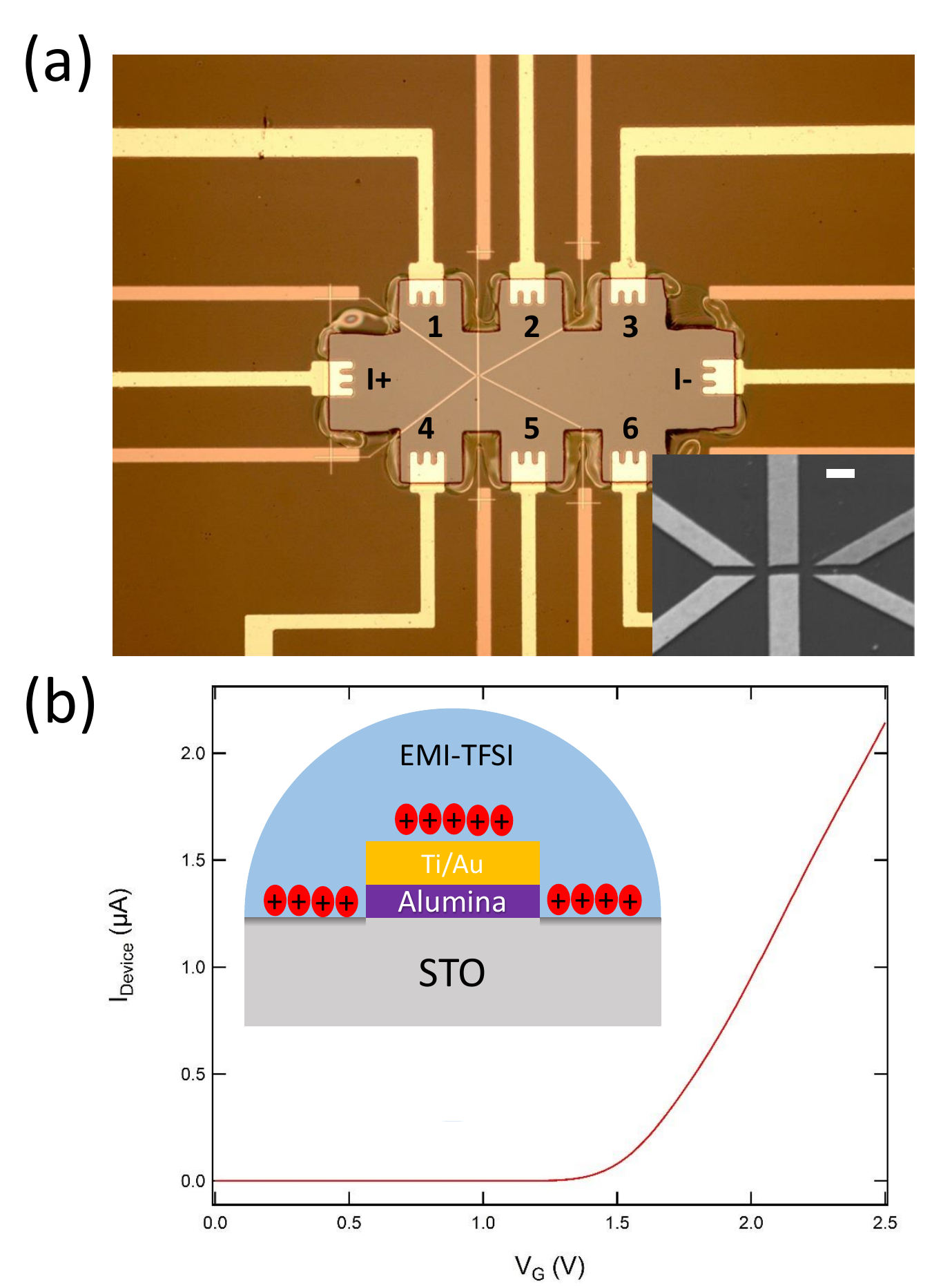}
\caption{(a) Optical image of active region of finished device; coplanar gate not shown. Inset: electron micrograph of nanoscale Hall bar region. White scale bar is 1 micron long. (b) Typical curve of device current versus coplanar gate voltage at room temperature, ramping the coplanar gate at 20 mV/second. Inset: Cross-section showing operation of a single side gate. No significant carrier accumulation occurs underneath the side gates.}
\end{figure} 

Prior to measurement, we apply a drop of the ionic liquid 1-ethyl-3-methylimidazolium bis(trifluoromethylamide), also known as EMI-TFSI (IoLiTec Inc., USA), that covers the larger Hall bar and the coplanar gate. With the sample in vacuum (<$10^{-2}$ mbar), we ramp up the voltage on the coplanar gate, drawing positive ions toward the unmasked STO regions. Crucially, mobile carriers are only induced in the unmasked regions; the photoresist and the gate metal sufficiently separate the ions from the STO so that negligible carrier density is accumulated underneath, shown schematically in the inset of Fig. 1b. Separate measurements confirm that there is no conduction underneath the nanoscale gates defining the smaller hall bar. In the measurements reported here, the voltage on the coplanar gate V$_G$ was ramped to +2.5 V at room temperature. Fig. 1b shows a device current I$_D$ as a function of V$_G$. This curve was measured with voltage bias between contacts 3 and 6, as the coplanar gate voltage was ramped at 20 mV/s at room temperature. The sample started to conduct when $\sim$1.2 V  was applied to the coplanar gate. Once the gate voltage reached +2.5 V, the sample, already in a dilution refrigerator, was rapidly cooled to a base temperature of 12 mK. The leakage current to the coplanar gate dropped below the measurement limit as the sample cooled through the glass transition temperature of the ionic liquid, about 175 K. The bulk region of the device, still measured between contacts 3 and 6, went superconducting at $T_c$= 220 mK.

At base temperature, we measured both the longitudinal and the transverse magnetoresistance of the device using standard lock-in techniques at 97 Hz. These measurements were all made in a 4-wire current-biased configuration, with current sourced between the ``I+'' and ``I-'' contacts (Fig. 1a). Longitudinal magnetoresistance $R_{xx}$ of the nanoscale Hall bar is measured using contacts 1 and 2, and transverse magnetoresistance $R_{xy}$ is measured using 1 and 4. We simultaneously monitor the transverse magnetoresistance of the bulk STO region by measuring the voltage drop between 3 and 6. The Hall density is $n=2 \times 10^{13}/cm^2$ inside the nanoscale Hall bar, and $n=4 \times 10^{13}/cm^2$ outside. The observed factor of 2 difference in density contrasts with previous work on LAO/STO microstructures, in which processing and confinement have little effect on the induced carrier density\cite{stornaiuolo_-plane_2012}. We speculate that the carrier accumulation due to the ions is weaker near the nanoscale gates, since the nearby grounded metal draws electric field lines away from the 2DES. The nanoscale Hall bar is superconducting at our measurement temperature of 13 mK, with a critical field $H_{c2} = 15 mT$, and a critical current density $J_c = 20 \mu A/cm$, comparable to those of Hall bars tens of microns wide with this carrier density.

Fig. 2a shows the magnetoconductance $G \equiv 1/R_{xx}$ at fields greater than $H_{c2}$ and temperatures between 12 mK and 520 mK. Universal conductance fluctuations are visible: aperiodic, reproducible fluctuations as a function of magnetic field. The fluctuations are suppressed by increasing temperature, but persist above 0.5 K. Following previous work on  UCF in LAO/STO heterostructures\cite{rakhmilevitch_phase_2010}, we assume (and then verify self-consistently) that the phase coherence length $ l_\phi \sim l_T \equiv \sqrt{hD/k_BT}$, where $l_T$ is the thermal length, $D$ is the diffusion constant and $T$ is the temperature. For our sample $l_T$ ranges from $\sim$300 nm at 520 mK to $\sim$2$\mu$m  at 12 mK. Under this assumption, we can estimate the dephasing rate $\tau_{\phi}^{-1}$ from the magnetic correlation field $B_{C}$ of the UCF:
\begin{equation}
\label{eqn:rate}
\tau_{\phi}^{-1}=\frac{{e}}{h}B_{C}D.
\end{equation}
Here $B_{C}$ is calculated from the conductance correlation
function
\begin{equation}
C(\Delta B) = \langle\delta g(B)\delta g(B+\Delta B)\rangle
\end{equation} using the condition $C(B_{C})=\frac{{1}}{2}C(\Delta B=0)$. We observe no significant dependence of $C(\Delta B)$ on $B$ up to 8 T, indicating no significant dependence of dephasing on magnetic field within our experimental range.

The calculated dephasing rate is plotted in Fig. 2b for various temperatures between 12 and 520 mK. 
We extract a dephasing rate of $7.33 \times 10^9$ Hz at 12 mK, corresponding to a phase coherence length $l_{\phi}=\sqrt{{Dt_{\phi}}} = 280$ nm. This calculation is self-consistent, as the calculated $l_{\phi}$ is within an order of magnitude of $l_T$. The dephasing rate rises linearly with temperature, as expected for electron-electron interactions in a two-dimensional disordered conductor\cite{imry_introduction_2009}. Interestingly, the dephasing rate clearly does not extrapolate to zero at zero temperature, suggesting a temperature-independent source of dephasing in our device, possibly local magnetic moments. Previous studies of EDLT-STO systems have shown Kondo effect at n > $6*10^{13}$ with $T_K > 10$ K \cite{lee_electrolyte_2011, li_role_2012}. At our lower carrier density, we might expect much lower $T_K$, so our entire experimental temperature range could be $\gg T_K$, a regime in which magnetic impurities are predicted \cite{vavilov_conductance_2003, zarand_theory_2004} and experimentally demonstrated \cite{micklitz_magnetic_2007, lundeberg_defect-mediated_2013} to yield constant-in-temperature dephasing.

\begin{figure}
\label{fig:temperature}
\includegraphics{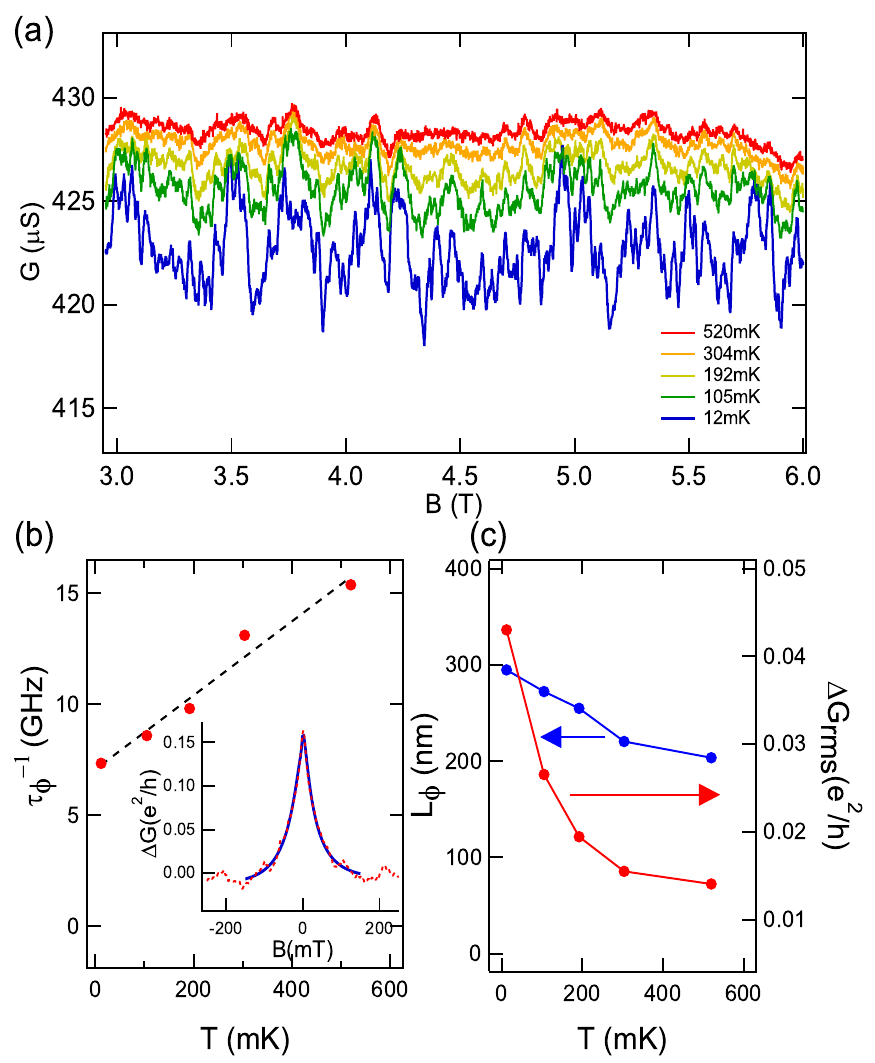}
\caption{(a) Device magnetoconductance $G$ over the temperature range 12-520 mK at zero side gate voltage. The reproducible universal conductance fluctuations in $G$ maintain shape but decrease in amplitude with increasing temperature. (b) Dephasing rate versus temperature with linear fit. Inset: $\Delta G(B)$ at T=520 mK, with fit to Eq. 3. (c) Phase coherence length and root mean square amplitude of conductance fluctuations versus temperature.}
\end{figure} 

\begin{figure}[b!]
\label{fig:gates}
\includegraphics{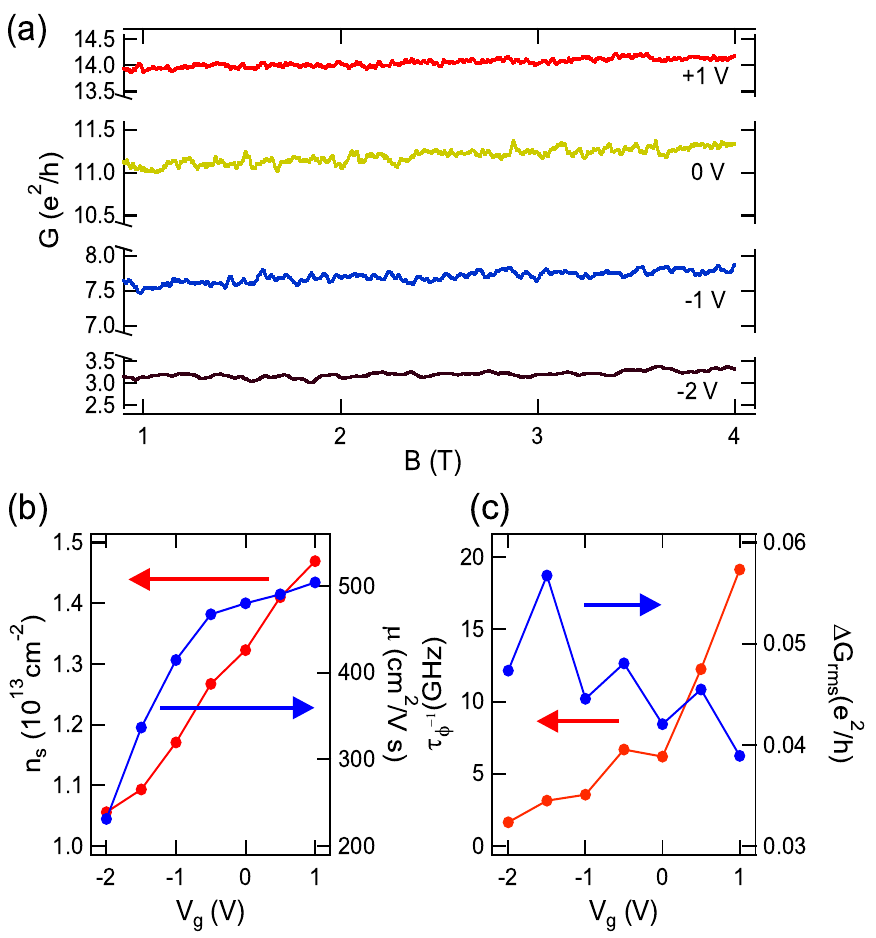}
\caption{(a) Magnetoconductance $G$ for different side gate voltages between -2 V and 1 V. Universal conductance fluctuations are clearly visible, but the details change between gate voltages as the disorder landscape evolves. (b) Electron density, extracted from the Hall slope, and mobility, extracted from the normal-state average conductance, as a function of V$_g$. (c) Dephasing rate and root mean square conductance fluctuation amplitude as a function of V$_g$.}
\end{figure}

The rms value of the fluctuations $\Delta G_{rms}$ is plotted over our temperature range in Fig. 2c. If the phase coherence length is larger than the sample width $w$ and length $l$, one expects $\Delta G_{rms} = \frac{e^{2}}{h}$ at $T=0$, and a decay in amplitude at higher temperatures as a result of thermal blurring. For $l_\phi$ less than the sample size, one expects $\Delta G_{rms} = \sqrt{\frac{l_{\phi}^{2}}{wl}}\frac{e^2}{h}$ due to incoherent addition of fluctuations at length scales greater than $l_\phi$. The fluctuations at our lowest temperature of 12 mK have an rms amplitude $\sim \frac{{1}}{20}\frac{{e^{2}}}{h}$, somewhat smaller than the $\sim \frac{1}{4}\frac{e^2}{h}$ we expect from the lithographic dimensions of our device. We do not have a clear reason for this discrepancy. Typically, dephasing estimates from UCF can be compared to estimates from an analysis of the weak antilocalization correction to $R_{xx}(B=0)$; in our device, the superconducting phase at zero field only allows for this analysis at T=520 mK, well above $T_c$. Previous calculations for weak localization in the presence of spin-orbit coupling \cite{hikami_spin-orbit_1980, rakhmilevitch_phase_2010,lee_electrolyte_2011} predict a weak antilocalization contribution to the conductance equal to: 

\begin{equation}
\label{eqn:weakloc}
\begin{split}
\Delta G(B)=\frac{e^2}{\pi h}(\Psi(\frac{1}{2}+\frac{H_1}{H})-\Psi(\frac{1}{2}+\frac{H_2}{H})+\\
\frac{1}{2}\Psi(\frac{1}{2}+\frac{H_3}{H})-\frac{1}{2}\Psi(\frac{1}{2}+\frac{H_4}{H})),
\end{split}
\end{equation} where $\Psi$ is the digamma function, $H_1=H_\phi+H_{so}^x+H_{so}^y+H_{so}^z$, $H_2=H_e$, $H_3=H_\phi+2 H_{so}^x+H_{so}^y$, $H_4=H_\phi$, and the characteristic field $H=\frac{h}{8 \pi e l^2}$. This equation can be simplified by assuming an isotropic spin-orbit coupling, so $H_1$=$H_3$. The fit of Eqn. 3 to our low-field magnetoconductance data at T=520 mK is shown in the inset of Fig. 2b. From the fit we extract $H_\phi=3.3\pm0.3$ mT, corresponding to $l_\phi=220\pm10$ nm, in close agreement with $l_\phi=210\pm10$ nm extracted from the correlation field of the UCF.

We next examine dephasing as a function of voltage on the nanopatterned gates. Although the ions in the ionic liquid are frozen in place, applying a voltage V$_g$ to the nanoscale gates can accumulate or deplete nearby carriers by conventional electric field effect. Equal V$_g$ is applied to all six of the gates at 12 mK. The normal state conductance changes by a factor of 4.5 as V$_g$ is varied from -2 V to 1 V (Fig. 3a). As gate voltage is made increasingly positive, electron density increases by a factor of 1.5 and mobility increases by a factor of 2, from $250$ cm$^{2}$/Vs to over $500$ cm$^{2}$/Vs (Fig. 3b). These numbers, obtained from Hall and longitudinal transport measurements, are approximate, since density and disorder must vary spatially given the gate geometry. We observe hysteresis in the gate sweeps, similar to the back gate hysteresis reported in ref.\cite{stornaiuolo_-plane_2012}.

Applying a voltage to the side gates changes the disorder landscape, altering the detailed fluctuations observed, but UCF remain at all gate voltages. The dephasing rate as a function of $V_g$ is calculated using Eq. \ref{eqn:rate}, and is shown in Fig. 3c, along with the rms conductance fluctuation amplitude.

The dephasing rate generally increases as carriers are added, but increases most dramatically between 0 and 1 V. There are several possible explanations for this behavior. For instance, application of a positive $V_g$ may accumulate carriers in previously depleted regions around the perimeter of the gate electrodes. Perhaps scatterers are enriched in this area compared to the inner channel of the Hall bar, due to the gate lithography. This would introduce a new dephasing mechanism for positive gate voltages, causing a sharp increase in dephasing rate. Alternatively, the system could be near a dimensionality transition: at 12 mK, $l_\phi \approx w$, the device width. If the system is approaching a regime where it is no longer two-dimensional -- strictly, $l_\phi \ll w$ --  the slight changes in channel width due to the changing side gate voltage could have a larger effect than expected. 

In summary, we have measured the transport properties of a gate-defined nanoscale Hall bar of strontium titanate, observing universal conductance fluctuations. From the UCF we extract the electronic dephasing rate as a function of temperature and side gate voltage. The linear temperature scaling of the dephasing rate suggests electron-electron interactions as a primary source of dephasing; the fact that the dephasing rate extrapolates to a nonzero value at $T=0$ indicates an additional temperature-independent source of dephasing, such as interaction with localized, unscreened magnetic moments. We also have demonstrated the ability to tune the carrier density, mobility, and dephasing rate with side gates. Our technique for nanoscale confinement and manipulation of STO 2DESs opens the door to future studies of the exotic superconducting and magnetic properties of this system.

\end{document}